# A Digital Twin Design Methodology for Control, Simulation, and Monitoring of Fluidic Circuits

Veyis Gunes

*Abstract*—This article proposes a methodology for the design of digital twins applicable to various systems (pneumatic, hydraulic, electrical/electronic circuits). The methodology allows representing the operation of these systems through an active digital twin, enabling the simulation, control, and monitoring through a more suitable and powerful computer-aided design. Furthermore, this methodology enables the detection of a system's actions on its own inputs (for instance, in pneumatics: backflow of gases trapped in parts of a fluidic system into its own inputs). During the simulation or monitoring phase, the approach also facilitates real-time diagnosis of the physical system. The outputs of the controlled physical system or its digital twin, do not depend only on the current inputs but also on the history of the inputs and the history of internal states and variables. In other words, the underlying sequential behavior logic has a memory while an combinational logic approach alone has not. These modeling capabilities can contribute to the digital transformation of the design, control, simulation, and manufacturing systems.

*Index Terms*— Control, Digital, Fluidics, Modeling, Twin.

## 1. Introduction[*]

Digital combinational circuits, by nature, operate in parallel, like fluidic circuits such as binary fluidic circuits (pneumatics, hydraulics) and circuits with electronic charges, i.e., electronic/electrical circuits. However, combinational logic (hardware or software) have a unidirectional operation and a time-independent behavior, which does not allow a logic and dynamic modeling (i.e., a behavior logic modeling) of a fluidic or distribution system and more generally, a physical system. We propose a simple, yet powerful, design methodology applicable to various domains. It allows the modeling of these systems through an active digital representation (i.e., a Digital Twin), and enables a more powerful computer-aided design, simulation, control, and monitoring. A Digital Twin (DT) can be defined as an active dynamic digital representation of a system, process or physical entity. Being active, any change (i.e., control), allowed on the representation, translates into a change of the corresponding variable on the physical entity or process. Other definitions, according to specific fields, are given in [1-4], and in [5-6], in the wide context of smart factories (and called "cyber-physical systems"). When a realistic rendering is not necessary, as in our examples, the DT can take the form of a schematic DT.

Applied to fluidic circuits, the proposed modeling involves static or quasi-static pressures of fluids, contained in tanks, cavities, or pipe sections, and subjected to overpressure or under pressure (thresholds allow the detection of such pressures). We address mainly the logical and dynamic modeling problem of these physical systems: how to design, simulate, visualize, control, and monitor complex circuits with pressure (and by analogy, potential) differences, on a computer? Our methodology allows detecting a system's actions on its own inputs (for instance, in pneumatics:

discharge into the inputs of gases trapped in parts of a system). Also, during simulation or operation, new (normal or abnormal) states could be detected and submitted to a human supervisor or an expert.

Reference [7] describes the possibility of designing classical logic functions within stationary fluidics but does not address potential application of these functions in a DT architecture. Others, such as [8] deal with the modeling of objects faults (in this case: rolling bearing). The patents [9] and [10] propose a representation of the control circuit, ignoring the underlying logic of the physical system (fluid propagation). Reference [11] introduces a method for tracking unidirectional moving objects but lacks a systematic/general modeling, while [12] constructs a dynamic schema (from event data streams) which, however, lacks a needed high reliability to control processes. Consequently, there isn't, yet, a unified definition of a DT in this field, nor in a cross-fields domain.

We propose a methodology establishing a series of simple logical equations, executed sequentially, to represent the behavior logic of a complex circuit in real-time, in order to design, simulate, control, and monitor it (examples: gases distributor, gases mixer, water distribution circuit). Two logical states are adopted: ambient pressure, which corresponds to a logical "0" and any overpressure/depression (a threshold can be set according to the application) corresponds to a logical "1". When the simulated system has to be materialized (implemented), a real physical system can be set up, and a threshold can be set for the sensors installed. Monitoring and control can then take the form of an active schematic DT. Thus, the DT becomes an active dynamic representation of the controlled system. It interacts with the system being controlled and animates itself as changes occur in the system (fluidic, electronic, electric, etc.).

This article introduces the problem set in section II. Then, the new methodology is proposed in section III. Section IV details the application of the methodology to three hypothetical examples and one real example. It concludes within section V.

[*]V. Gunes was with *Femto-ST*, ENSMM, Université de Franche-Comté, 25030 Besançon, France. He is now with the *Institut des Molécules et Matériaux du Mans*, Le Mans Université, 72085 Le Mans, France (e-mail: veyis.gunes@univ-lemans.fr).

Color versions of one or more of the figures in this article are available.

## 2. The problem set

This article deals mainly with the underlying behavior logic of a DT system that accurately mirrors the dynamics of a fluidic or electrical distribution system. This modeling should take into account the bidirectional flow scenarios and the challenges in representing a system faithfully with its behavior logic. More generally, these aims involve the modeling of physical systems, their simulation, and the design of Digital Twins. Our approach is based on a three dimensional DT system defined or used, for instance, in [2,13,14], where the "bidirectional data link" or "data" (third dimension) becomes, in our case, a whole underlying or behavioral model. Thus, our model is composed of: a Physical System/circuit/process (PS), a Virtual System/circuit/process (VS), and an Underlying Model (UM) which is regularly synchronized (updated), with both PS and VS (Fig. 1). The UM is also named, by Wang *et al.* [15], "the proxy model" and includes "State prediction" and "Fault diagnosis", as in our approach. Wang *et al.*[16] name it "a link module" and use an OPC UA server to model it. The hypothetical example of Figure 2 is used as an illustration of our UM example. In this definition, a DT system can be designed and simulated before the existence of the PS. Indeed, the "Activation" and the "Updates" from the PS can be disabled in design and simulation stages. Any active DT system should integrate at least one common parameter, used in the three dimensions (parts) of the DT system.

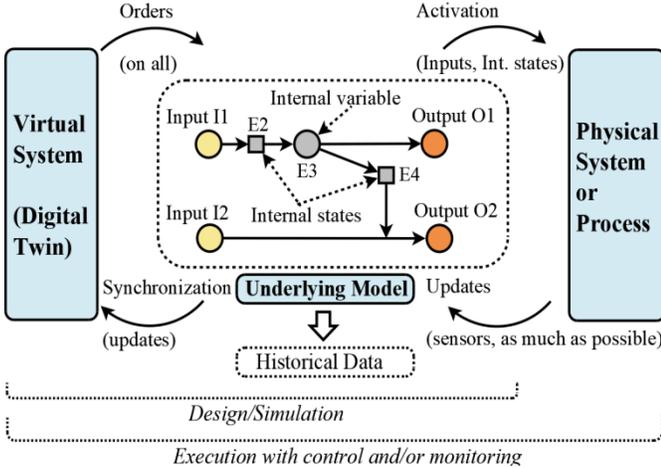

**Fig. 1.** A three dimensional DT system showing the interactions between its three parts: Virtual, Physical, and Underlying Model parts. This latter model is a Slave for the virtual part (Orders) and a Master for the physical part (Activation). Inputs I1 and I2 and outputs O1 and O2 represent the pipe sections E1, F1, S1, and S2 of Figure 2, respectively.

### 2.1. Basic example of fluid distribution circuit

Considering a distribution circuit with 2 inputs and 2 outputs, represented in Figure 2, we associate a Boolean state to each pipe. The two logical states are determined as explained in the introduction. The threshold value should be enough low to allow detection, even if a pipe fluid spreads in other pipes.

The solenoid valves (or pneumatic valves) $E_2$ and $E_4$ will be considered as *internal states* (IS) since they depend only on the controls and do not depend on the system behavior logic. $E_3$ will be referred to as an *internal variable* (IV). Its state depends on the history of the process/circuit (i.e., on its behavior logic), so on the history of inputs, outputs, other IVs, and the ISs. Since, in fact, the flows can become bidirectional, the pipes $E_1$, $E_3$, $F_1$, $S_1$, and $S_2$ should be considered as both inputs and outputs (in the UM).

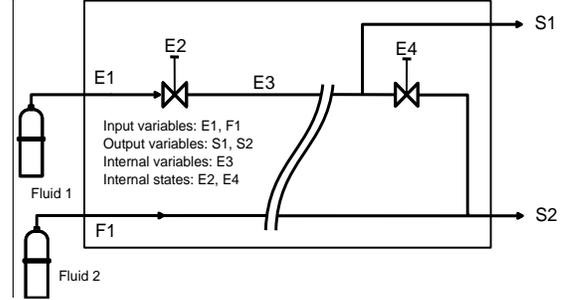

**Fig. 2.** A simple hypothetical example of a fluid distribution circuit (1) where $E_4$ has been added later in order to be able to possibly supply the second output ($S_2$), with the choice between fluid 1 or 2. The arrows denote the usual (normal) directions of flows. Each pipe or set of joined pipes is transformed in a Boolean input/output/internal variable. The double curves denote that the right components can be far from the components on the left.

In the UM, the main problems can thus be summarized in how to automatically predict the presence of fluids in $S_1$ or $S_2$ and the backflows on the inputs and/or *internal variables*, with or without the availability of feedbacks from optional sensors. Whenever sensors are available on the outputs, the IVs, and even on the ISs, their corresponding Boolean value should be used to update the UM. Thus, the experimental data is used to update the UM through the sensors. Also, this makes it possible to detect an abnormal situation (see section 4).

### 2.2. Limitation of a modeling by a combinatorial logic

If we consider that the flows are unidirectional (inputs to outputs), we can easily model this circuit as a digital system (in particular, $E_3$ would be considered here as an input, only).

By modeling each pipe section and valve as a logical input, a combinational system can be defined using Boolean algebra. We assume that all input variables and the IVs (here, $E_3$) are independent and that all output variables are independent from each other. This yields to:

$$S_1 = E_1.E_2.E_3 \tag{1}$$
$$S_2 = E_1.E_2.E_3.E_4 + F_1 \tag{2}$$

where we denote the conjunctive operator as "." and the disjunctive operator as "+". The following truth table is obtained (see Table 1):

TABLE 1
TRUTH TABLE OF CIRCUIT 1, CONSIDERED UNIDIRECTIONAL

| $F_1$ | $E_4$ | $E_3$ | $E_2$ | $E_1$ | $S_1$ | $S_2$ |
|---|---|---|---|---|---|---|
| 0 | 0 | 0 | 0 | 0 | 0 | 0 |
| 0 | 0 | 0 | 0 | 1 | 0 | 0 |
| 0 | 0 | 0 | 1 | 0 | 0 | 0 |
| 0 | 0 | 0 | 1 | 1 | 0 | 0 |
| .. | .. | .. | .. | .. | .. | .. |
| 0 | 1 | 1 | 1 | 0 | 0 | 0 |
| 0 | 1 | 1 | 1 | 1 | 1 | 1 |
| 1 | 0 | 0 | 0 | 0 | 0 | 1 |
| .. | .. | .. | .. | .. | .. | .. |
| 1 | 1 | 1 | 1 | 1 | 1 | 1 |

In digital circuits (electronics), an input cannot become an output because it is unidirectional. However, in fluidics, this is possible: a bidirectional flow is shown, as an example, in the shaded line, where there is a "backflow" of fluid in pipe $S_2$ into the input pipe $F_1$ ($F_1$ should switch to the state "ON"). So, the classical truth table is therefore not suitable for this modeling (e.g., a fluidic system) but it simply allows the study of a fluidic system for a unidirectional operation. Moreover, this approach does not allow considering internal pipes as both inputs and outputs, with the possibility of memorizing their previous state.

## 3. A methodology of logical and sequential modeling

The example introduced in Figure 2 includes 5 variables (inputs, ISs, and IVs), which already leads to $2^5$ (or 32) combinations. With more variables, truth tables become quickly too large for practical study, even with the application of simplification techniques. Furthermore, this modeling does not take into account previous states, neither the possible backflows of fluids (or feedbacks, in general) towards the inputs. The proposed methodology is based on causality (an effect cannot precede its cause). Initially, we establish the equations of the *internal variables* (i.e., pipes or cavities that can contain fluids or mixture of fluids) based on the *inputs*, only (step 1). Then, we establish the *output* equations, based on the *inputs* and *internal variables* (step 2). Finally, we determine the possible backflows or feedbacks towards the IVs and then, towards the inputs (steps 3 and 4, respectively). We will use a programming framework, such as the LabView Virtual Instrument concept, which can become a convenient way to implement the VS and the UM within the "front panel" and the "diagram" of LabView, respectively.

### 3.1. Identifying Inputs, Outputs, Internal Variables and States

Considering that all outputs and IVs are also inputs and all (real) inputs are also outputs (except where a one-way direction is installed: a check valve for fluidic circuits or a diode for electrical/electronic circuits), we can apply the rules:

*Dependence of "pipes" placed in series → Modeling by an AND function (represented ".").*
*Dependence of "pipes" joining in a node → Modeling by an OR function (represented by "+").*

$E_1$, $F_1$: inputs (and outputs, for backflows)
$E_2$, $E_4$: internal states or ISs (they are always inputs)
$E_3$: internal variable or IV (it can be an input and an output)
$S_1$, $S_2$: outputs (and inputs for backflows)

Note: $E_2$ and $E_4$ are actuators (here, fluidic circuit breakers) and they are not supposed to contain fluids. These ISs can be present in all equations and are derived from the control states of the VS and, if the PS allows a feedback from the ISs, from Boolean expressions with the corresponding sensors' values (example: Flow > 10 liters/hour?).

## 3.2. Establishing the equations in Boolean algebra

### 3.2.1 Inputs to IVs and outputs (usual direction of flows)

*Step 1*: Establishing IVs as functions of the inputs, by searching in all "branches" or directions. We have to stop if we reach an IV or an output (excluding terms with outputs in the equation) and add (OR function) the previous state of this variable (because a pipe may have been filled previously). Thus, we are taking into account the history of the circuit.

$$E_n = \sum_{i=1}^{I}(\prod_{j=1}^{J} P_{ij}) + E_{np} \qquad (3)$$

I is the number of branches, while J is the number of elements in series "conducting" to an input, in the $i^{th}$ branch. $P_{ij}$ is the state of the $i^{th}$ pipe of the $j^{th}$ branch whereas $E_{np}$ is the previous state of $E_n$. In the example:

$$E_3 = E_1 . E_2 + E_{3p} \qquad (4)$$

where p indicates the previous state of $E_3$.

*Step 2*: Establishing outputs as functions of the inputs and the IVs. We have to stop when we reach an IV or an input. We include the terms with these, in the corresponding equations and add (OR function) with the previous state of the output in this way:

$$S_n = \sum_{i=1}^{I}(\prod_{j=1}^{J} P_{ij}) + S_{np} \qquad (5)$$

where I is the number of branches, J the number of elements in series "conducting" to an IV or an input. Pij has been defined previously. $S_{np}$ is the previous state of $S_n$. In the example:

$$S_1 = E_3 + S_{1p} \qquad (6)$$
$$S_2 = F_1 + E_4 . E_3 + S_{2p} \qquad (7)$$

It is assumed that the consumption in S1 and S2 is low (quasi-static pressures after state changes) and does not "empty" or "consume" the IVs (fluids or, for electrical circuits, charges may be "trapped" inside).

### 3.2.2 Detection of backflows (for IVs and then inputs)

This detection involves flows from the outputs to IVs and then to the inputs: this is the direction of backflows or unusual flows, which also updates the UM by comparing it to the PS. We do not take into account previous states because, in this case, the outputs or IVs are assumed to be gradually consumed (or "discharged" for an electrical circuit).

In this case, we consider that:

- All inputs and IVs are also backflow outputs (except where a unidirectional flow is forced, with a check valve for fluidic circuits or a diode for electrical circuits).
- If we reach a circuit section (pipe) in contact with inputs (or IVs) and outputs, we only take the terms with these outputs because these outputs are the ones which can impose a backflow. This translates itself into a simplification rule.

*Step 3*: Establishing internal variables as a function of the outputs and IVs (backflow or feedback to IVs). We do not consider terms with inputs as their effects have already reached the outputs and IVs. We have to stop if we reach an IV or an output (included in the equation). We do not add the previous state because only the outputs and IVs should update the state of the calculated IV. The IVs can either be filled/charged by backflow or emptied/discharged because the connected outputs could be emptied/discharged. These operations can be done through orders from the VS (it's a "manual" update) or be updated automatically from the PS (after a first iteration), if sensors are installed over these outputs and IVs (see Fig. 1). In the example:

$$E_3 = S_1 + E_4.S_2 \qquad (8)$$

If this equation changes $E_3$ state to "1", it means that there is a backflow or propagation of an IV or an output (here, $S_1$ or $S_2$) to this IV. In this case, we assign the calculated value to $E_3$ only after a warning of a backflow to the user (the operator should be notified before assignment). Thus, if an IV or an output switches to "1", all IVs in "connection" with it can also switch to "1" in the following iterations, depending on the ISs.

Below, in the truth table of the modeled system (Table 2), (→) indicates changes compared to the unidirectional truth table presented above (its values taken as initial values), after the execution of the first 3 steps. These changes are due to the fact that all elements (except ISs) become inputs and outputs, depending on the step executed and, moreover, outputs and IVs can memorize their previous state.

The table shows that the constructed model is consistent with the behavior logic of the PS: the different steps produce the expected behaviors.

**Step 4**: Establishing the inputs as a function of internal variables and outputs (backflow or feedback to inputs). We have to stop if we reach an IV or an output (these terms are included). We do not add the previous state because only the outputs and IVs should update the state of the calculated input (IVs can be emptied/consumed or filled by backflow, and outputs can be emptied/consumed).

$$E_{1r} = E_2.E_3 \qquad (9)$$
$$F_{1r} = S_2 \qquad (10)$$

If $E_{1r}$ ="1" and $E_1$="0": a backflow warning is generated and $E_{1r}$ is assigned to $E_1$. These are the return values (backflows/feedbacks): if they are equal to a logical "0", we do not assign these values to the corresponding inputs because their states remain imposed by the (real) inputs (they can be seen as inexhaustible sources), hence the use of indices r for return flows.

This has been checked in the truth table (see Table 3). After the execution of step 4, backflow/feedback detections are underlined in the columns $E_{1r}$ and $F_{1r}$, in the table, where the truth table is the evolution of the previous table and is consistent with the behavior logic.

### TABLE 2
#### TRUTH TABLE WITH BACKFLOWS ON INTERNAL VARIABLES

| $F_1$ | $E_4$ | $E_3$ | $E_2$ | $E_1$ | $S_1$ | $S_2$ | From step: |
|---|---|---|---|---|---|---|---|
| 0 | 0 | 0 | 0 | 0 | 0 | 0 | |
| 0 | 0 | 0 | 0 | 1 | 0 | 0 | |
| 0 | 0 | 0 | 1 | 0 | 0 | 0 | |
| 0 | 0 | 0→1 | 1 | 1 | 0→1 | 0 | Step 1, 2 |
| 0 | 0 | 1 | 0 | 0 | 0→1 | 0 | Step 2 |
| 0 | 0 | 1 | 0 | 1 | 0→1 | 0 | Step 2 |
| 0 | 0 | 1 | 1 | 0 | 0→1 | 0 | Step 2 |
| 0 | 0 | 1 | 1 | 1 | 1 | 0 | |
| 0 | 1 | 0 | 0 | 0 | 0 | 0 | |
| 0 | 1 | 0 | 0 | 1 | 0 | 0 | |
| 0 | 1 | 0 | 1 | 0 | 0 | 0 | |
| 0 | 1 | 0→1 | 1 | 1 | 0→1 | 0→1 | Step 1, 2 |
| 0 | 1 | 1 | 0 | 0 | 0→1 | 0→1 | Step 2 |
| 0 | 1 | 1 | 0 | 1 | 0→1 | 0→1 | Step 2 |
| 0 | 1 | 1 | 1 | 0 | 0→1 | 0→1 | Step 2 |
| 0 | 1 | 1 | 1 | 1 | 1 | 1 | |
| 1 | 0 | 0 | 0 | 0 | 0 | 1 | |
| 1 | 0 | 0 | 0 | 1 | 0 | 1 | |
| 1 | 0 | 0 | 1 | 0 | 0 | 1 | |
| 1 | 0 | 0→1 | 1 | 1 | 0→1 | 1 | Step 1, 2 |
| 1 | 0 | 1 | 0 | 0 | 0→1 | 1 | Step 2 |
| 1 | 0 | 1 | 0 | 1 | 0→1 | 1 | Step 2 |
| 1 | 0 | 1 | 1 | 0 | 0→1 | 1 | Step 2 |
| 1 | 0 | 1 | 1 | 1 | 1 | 1 | |
| 1 | 1 | **0→1!** | 0 | 0 | 0 | 1 | **Step 3** |
| 1 | 1 | **0→1!** | 0 | 1 | 0 | 1 | **Step 3** |
| 1 | 1 | **0→1!** | 1 | 0 | 0 | 1 | **Step 3** |
| 1 | 1 | 0→1 | 1 | 1 | 0→1 | 1 | Step 1, 2 |
| 1 | 1 | 1 | 0 | 0 | 0→1 | 1 | Step 2 |
| 1 | 1 | 1 | 0 | 1 | 0→1 | 1 | Step 2 |
| 1 | 1 | 1 | 1 | 0 | 0→1 | 1 | Step 2 |
| 1 | 1 | 1 | 1 | 1 | 1 | 1 | |

Note: (1!) indicates a risk of backflow/feedback on internal pipes (or IVs, after the execution of steps 1, 2, and 3) which, in turn, can backflow into other parts. In these cases, it is necessary to warn the operator and then assign the relevant variables.

### TABLE 3
#### TRUTH TABLE WITH BACKFLOW DETECTION ON INPUTS

| $F_{1r}$ | $E_4$ | $E_3$ | $E_2$ | $E_{1r}$ | $S_1$ | $S_2$ |
|---|---|---|---|---|---|---|
| 0 | 0 | 0 | 0 | 0 | 0 | 0 |
| 0 | 0 | 0 | 0 | 1 | 0 | 0 |
| 0 | 0 | 0 | 1 | 0 | 0 | 0 |
| 0 | 0 | 0→1 | 1 | 1 | 0→1 | 0 |
| 0 | 0 | 1 | 0 | 0 | 0→1 | 0 |
| 0 | 0 | 1 | 0 | 1 | 0→1 | 0 |
| 0 | 0 | 1 | 1 | **0→1!** | 0→1 | 0 |
| 0 | 0 | 1 | 1 | 1 | 1 | 0 |
| 0 | 1 | 0 | 0 | 0 | 0 | 0 |
| 0 | 1 | 0 | 0 | 1 | 0 | 0 |
| 0 | 1 | 0 | 1 | 0 | 0 | 0 |
| 0 | 1 | 0→1 | 1 | 1 | 0→1 | 0→1 |
| **0→1!** | 1 | 1 | 0 | 0 | 0→1 | 0→1 |
| **0→1!** | 1 | 1 | 0 | 1 | 0→1 | 0→1 |
| **0→1!** | 1 | 1 | 1 | **0→1!** | 0→1 | 0→1 |
| **0→1!** | 1 | 1 | 1 | 1 | 1 | 1 |
| 1 | 0 | 0 | 0 | 0 | 0 | 1 |
| 1 | 0 | 0 | 0 | 1 | 0 | 1 |
| 1 | 0 | 0 | 1 | 0 | 0 | 1 |
| 1 | 0 | 0→1 | 1 | 1 | 0→1 | 1 |
| 1 | 0 | 1 | 0 | 0 | 0→1 | 1 |
| 1 | 0 | 1 | 0 | 1 | 0→1 | 1 |
| 1 | 0 | 1 | 1 | **0→1!** | 0→1 | 1 |
| 1 | 0 | 1 | 1 | 1 | 1 | 1 |
| 1 | 1 | **0→1!** | 0 | 0 | 0 | 1 |
| 1 | 1 | **0→1!** | 0 | 1 | 0 | 1 |
| 1 | 1 | **0→1!** | 1 | 0 | 0 | 1 |
| 1 | 1 | 0→1 | 1 | 1 | 0→1 | 1 |
| 1 | 1 | 1 | 0 | 0 | 0→1 | 1 |
| 1 | 1 | 1 | 0 | 1 | 0→1 | 1 |
| 1 | 1 | 1 | 1 | **0→1!** | 0→1 | 1 |
| 1 | 1 | 1 | 1 | 1 | 1 | 1 |

Note: (1!) indicates a risk of backflow/feedback towards the inputs or the IVs (pipes or lines). Backflows to the inputs are underlined.

### 3.3. The different steps of the process (Algorithm)

The process of identifying, designing, and the execution of the DT by the proposed methodology is described below, in algorithm 1:

**Algorithm 1** Generation and execution of the UM/DT.

- Identify and create the Inputs, Outputs, IVs, and ISs of the UM and DT.
- Establish equations defined in **steps 1 to 4** (this can be computerized through the construction of an directed graph).
- Initialize the VS (Inputs, ISs, and IVs).
- **Repeat** infinitely (sequential execution):
  1. Assign Inputs, ISs, IVs, and Outputs of the VS to the UM.
  2. Execute the logical equation for each IV (step 1).
  3. Execute the logical equation for each output (step 2).
  4. Assign the inputs and the ISs in the PS (when the PS is available).
  5. Pause (typically 100 ms): according to reaction time of the process/system (PS available).
  6. Update Outputs and IVs in the UM, if there are sensors on them (PS available). Otherwise, their states should be changed in the VS (by a click on the concerned pipe) into their current states (e.g., if they are emptied/discharged "manually").
  7. Execute the logical equation for each IV (backflows/feedbacks to IVs). If a result is "ON" and the current state is "OFF", warn of a backflow on the related IV (step 3) and assign the result in the UM.
  8. Execute the equation for each input (step 4) to detect backflows on the inputs and process as described in step 4.
  9. Update the VS according to the UM (outputs and IVs).
- **End repeat**.

### 3.4. Advantages of the methodology

The main advantages of this methodology are:

- Representation of the state of a system through an active dynamic representation (or active VS), in real-time. This latter is reached if the "Pause" value is higher than any involved effectors response time within the PS.
- Systematic (and programmable) application: prior to determining the equations, a schematic input system can allow the construction of a directed graph by attributing the inputs, outputs, IVs, and ISs. The graph could enable the establishment of equations (steps 1 to 4), thereby simplifying the synthesis of the set of logical equations (in the UM).
- Possible implementation in software or hardware. In the latter case (with logic gates and flip-flops, the method leads to simple equations but it wouldn't allow an easy simulation.
- Detection of backflows: in design/simulation, it is possible to determine precautions to take (deactivations of certain possibilities). In operation/monitoring mode, we can warn operators and trigger an alert.
- Possibility of diagnosis (see next section).

### 4. Application in Design, Simulation, Monitoring, and Control Systems

It is possible to store, in the form of a binary word, all situations considered "normal" by an expert. The binary word will consist of as many bits as there are variables (inputs, outputs, ISs, and IVs), with each variable represented by a bit. At each change in this word, we can compare the new word to the set of existing words (validated/added by an expert), in the Historical Data (HD). If this word is new, an alert can be triggered, and/or an expert's intervention/validation can be requested. A continuous diagnosis also allows the detection of potential leaks or blockages, because such situations would generate new words. Each new word can be added to the HD with a label describing it as "normal", "unseen", "backflow", "leakage", "blockage", etc., Additionally, as suggested in [12], sensors instrumenting the physical process, would allow the exact location of these problems. Once the HD is rich enough, supervised or unsupervised machine learning algorithms can be implemented on the HD for a higher level of diagnostics. The main application fields of the methodology are:

### 4.1. In Fluidics (Industrial Processes, Water Supply)

This methodology is well adapted for:

- Modeling, design, and simulation of complex fluidic systems such as distributors, mixers, etc.
- Creation of Digital Twins enabling real-time control of a fluidic system.
- Detection of backflows towards inputs, tanks, and internal pipes or lines (to avoid pollution or to detect leaks).
- Detection of leaks and blockages in the pipes (whenever sensors are installed on the PS and their values compared to the corresponding values in the UM).

### 4.2. In Electricity/Electronics (direct current)

In certain conditions, especially when the conductors are isolated, an electron flow can be seen as a particular case of fluid. Then, the various potentials in such a circuit are analogous to fluid pressures.

- Modeling, designing, and simulating of complex networks (e.g., voltage multiplier),
- Creating Digital Twins controlling or simulating complex electrical circuits in real-time,
- Detection of over-voltages or unwanted voltages on inputs (e.g., electrocution risks, detection of outputs applied to inputs).

### 4.3. Example of a VS/DT simulation (case in Figure 2)

Figure 3 shows an example of a virtual process (or VS) in which blue color identifies the presence of fluids and red color identifies a backflow problem. The VS (i.e., DT) and the UM are represented/implemented in the front panel and the block diagram of the LabView "virtual instrument", respectively. We can click on the inputs or the ISs to change their state and on the outputs or the IVs to simulate the consumption or the emptying. Whenever some sensors measuring the real values (IVs, and outputs) are available, these values will allow the update of the model (UM) and then, the synchronization of the VS (see Figure 1). The available sensors could be, in this case, associated with one or more pressure thresholds and provide logical signals.

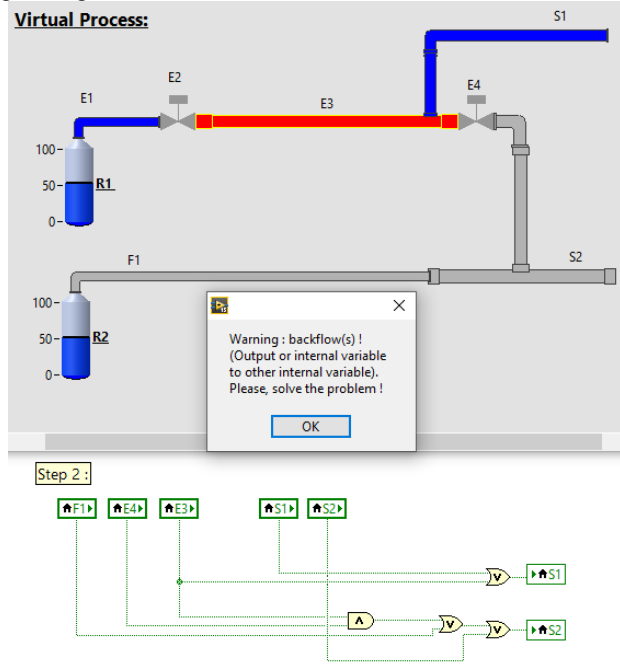

**Fig. 3.** Example circuit (1) with 2 gas sources (R1, R2) and 2 outputs (S1, S2). When a pipe is emptied (no overpressure), it has to be set to OFF on the VS, just by clicking on it. Here, $E_1$ and $E_2$ were previously activated, filling $E_3$ and $S_1$. If $E_3$ is emptied (forced to "OFF"), there is a backflow from $S_1$ into $E_3$, which becomes red/blinking to indicate a backflow. If the message is acknowledged (OK) and $S_1$ is emptied (i.e., set to OFF), $E_3$ returns to a grey color. The lower part shows the implementation of step 2 (only) in the block diagram, as an example.

### 4.4. Application to a more complex fluidic system – Example 2

The application of Figure 4 has been designed using the described methodology and implemented using the LabVIEW programming environment. The different steps are:

• **Initial step - Identification**

$E_1$, $F_1$, $G_1$: input (and output) variables

$E_2$, $E_4$, $E_5$, $F_2$, $G_2$, $H_2$: internal states (inputs)

$E_3$, $H_1$: internal variables (input and output)

$S_1$, $S_2$: output (and input) variables

Note: $E_2$, $E_4$, $E_5$, $F_2$, $G_2$, $H_2$ are actuators (or ISs) and are not supposed to contain fluids.

• **Step 1**

$$E_3 = E_1.E_2 + E_{3p} \tag{11}$$

$$H_1 = F_1.F_2 + G_1.G_2 + H_{1p} \tag{12}$$

• **Step 2**

$$S_1 = E_5.E_3 + S_{1p} \tag{13}$$

$$S_2 = H_2.H_1 + S_{2p} \tag{14}$$

• **Step 3**

$$E_{3r} = E_5.S_1 + E_4.H_{1p} \tag{15}$$

$$H_{1r} = H_2.S_2 + E_4.E_{3p} \tag{16}$$

(p indicates the previous iteration state)

• **Step 4**

$$E_{1r} = E_2.E_3 \tag{17}$$

$$F_{1r} = F_2.H_1 \tag{18}$$

$$G_{1r} = G_2.("0") = "0" \tag{19}$$

no possible backflows (cv: check valve)

• **Simulation of the VS/DT implemented with LabView**

In the situation presented in Figure 4, we have filled the pipes $S_1$, $E_3$, and $H_1$ through $E_2$, $E_5$ and $E_4$ (with $H_2$ closed), then closed $E_2$, $E_5$, and $E_4$, and emptied $E_3$ (then in grey or OFF). When we reopen $E_4$, a backflow on $E_3$ is then detected and $E_3$ is shown, in the figure, in red and a blinking state. We also opened $G_2$ and $F_2$, so this latter will cause a backflow warning on $F_1$, after acknowledging the current warning. There is no backflow on $G_1$ because the check valve (cv) prevents a backflow into $G_1$.

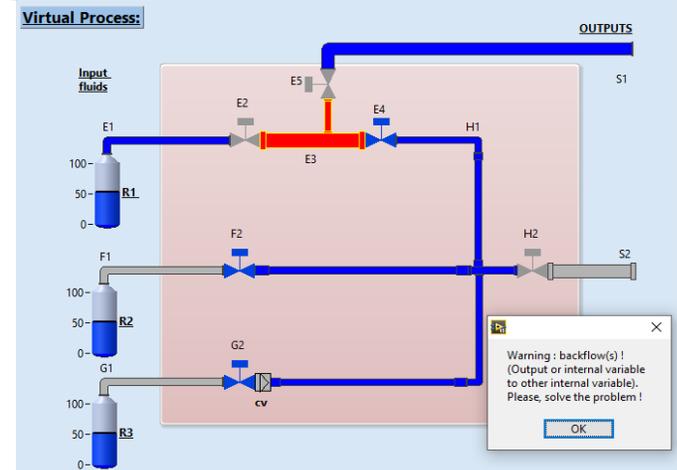

**Fig. 4.** Example of circuit (2). There is a warning of a backflow from $H_1$ into $E_3$ (which is in red and blinking). If we click "OK", $E_3$ turns blue, and a second backflow will be announced on the input $F_1$ since we also opened $F_2$.

### 4.5. Application to an electrical/electronic circuit – (equivalent to Cicuit 2)

Below, Figure 5 shows how the methodology can be applied to an electrical/electronic circuit. The circuit operates similarly to Circuit 2 (which is for fluid distribution). The conductors are analogous to the pipes of the previous circuits. Thus, the presence of pressure is replaced by the presence of voltage. The methodology could enable the design, simulation, control, and monitoring of voltage multipliers or adders.

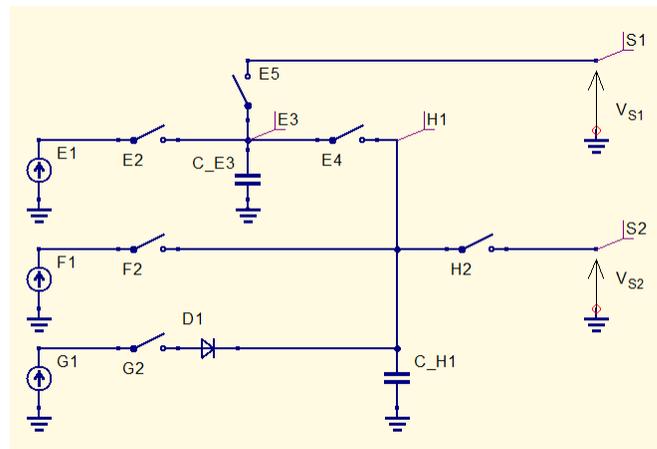

**Fig. 5.** Example of electrical/electronic circuit which is equivalent to the fluidic circuit of Figure 4. The two capacitors constitute IVs (in this case, they may store electrical energy, instead of fluid, as in the previous example).

If we apply the methodology, after charging capacitors C_E3, C_H1 through E$_2$ and E$_4$ (with E$_5$ and H$_2$ opened), and then opening E$_4$ and discharging C_E3 (for instance, by supplying energy on output S$_1$ through E$_5$), we would induce warnings similar to the previous circuit. For instance, closing E$_4$ causes a charge of capacitor C_E3, so a backflow on an IV. Then, if we close F$_2$, we trigger a "presence of voltage" warning on F$_1$ even if the voltage source connected to F$_1$ is OFF. Consequently, F$_1$ and E$_3$ can be warned for backflow (for example, by making them blink, as in the previous circuit). There is no backflow on G$_1$ because this input is protected by diode D1 which imposes a unidirectional flow.

## 4.6. Integration in a complex impedance spectrometry system

We have applied the methodology in a Complex Impedance Spectrometer (CIS) system. The system is currently under development in our laboratory. It is composed of a PC that controls a Middle Frequency Impedance Analyzer (MFIA, determining the impedance of a sample, connected through a USB port), a furnace capable of heating a measuring cell up to 1000°C (connected through an RS232 port, equipped with a Eurotherm 3216 temperature controller), and a gas mixing system (see Figure 6). Only the DT tab (named "Setup") of our specific CIS software is shown. The gas evaporator (Heater) for this mixing system is optional. However, in its absence, we can still interact/click on the PC_H signal on the DT to check and use the remaining circuit without being disturbed by its absence. This DT allows a monitoring of the fluidic circuit and a monitoring of the equipment components activity. Thus, it takes also into account the behavior of our specific CIS software. In the future, we can replace the manual valves (v_man1, v_man2) with controlled valves so that we can control a selection or mixing of gases (directly or in a recipe defined in the "Configuration" tab) through the DT. If we install also sensors in the pipes we would be able to detect leaks and blockages of gases.

In this application, we also use the proposed methodology for the signal/control flow circuit. For this circuit, we apply only relevant steps: Z_at_T output is a logical signal, so it deals only with unidirectional flow (the answers of the impedance analyzer and the furnace are only used as "sensors" to validate the correct connection and communication). Also, we do not need to "add" previous values, as this data flow is not stored and there are no backflow problems. The G_at_T signal plays a role in both circuits. It is activated when the furnace has reached its setpoint temperature and when the gas (mixed or not) sweeps the cell. The UM steps that apply to this system are (variables are Booleans, except noted otherwise):

- **Initial step - Identification**
  Fluid Circuit: (v_man1, v_man2 not included in the study)
  Input Variables: Gas 1, Gas 2
  Internal States: PC_H, PC_TC, Trd (see notes, below)
  Internal Variables: Gm, Gh
  Output Variables: G_at_T  (Gas at setpoint temperature)

*Signal/Control Circuit (in italic):*
*Input Variables: PC_CIS, SP Furnace (Real)*
*Internal states:*
  *G_at_T,*
  *SWP_on (measurement, sweeping in frequency)*
*Internal Variables: MFIA_out*
*Output Var.: Z_at_T (CIS results in progress), Target (Real)*

Notes:
- SWP_on is a signal indicating that a frequency sweep is in progress (it is active after the diffusion delay in the sample/cell).
- PC_CIS: a logical state of the communication with the MFIA (updates every second).
- PC_TC: state of the communication with the temperature controller (furnace).
- PC_H: state of the communication with the evaporator (Heater).
- MFIA_out is the state of the applied AC signal.
- Trd: signal indicating that "Temperature" has reached the setpoint temperature ("SP Furnace", in Figure 6).

- **Step 1**

$$Gh = Gas\ 2 \cdot PC\_H + Gh_p \qquad (20)$$

$$Gm = Gas\ 1 + Gh + Gm_p \qquad (21)$$

$$MFIA\_out = SWP\_on \cdot PC\_CIS \qquad (22)$$

- **Step 2**

$$G\_at\_T = Trd \cdot PC\_TC \cdot Gm + G\_at\_T_p \qquad (23)$$
(if G_at_T = "ON", the color of the furnace changes to brown and "ready" is displayed)

$$Z\_at\_T = G\_at\_T \cdot MFIA\_out \qquad (24)$$
(states of "cell" and "MFIA_in" are copies of Z_at_T)

$$Target = SP\ Furnace \qquad (25)$$

- **Step 3**

$$Gm = PC\_TC \cdot Trd \cdot G\_at\_T + Gh \qquad (26)$$
(if Gm="ON": warning and then Gm="ON")

$$Gh = Gm \qquad (27)$$

- **Step 4**

$$Gas\ 1_r = Gh + Gm \qquad (28)$$

$$Gas\ 2_r = PC\_H \cdot Gh \qquad (29)$$

$$SP\ Furnace = Target \qquad (30)$$

In the figure, the furnace has reached its setpoint (here, "Ready" at 100°C). Also, the pipes or conductors Gas1, Gm, PC_CIS, and PC_TC are in their active colors, but the "Cell" is not yet ready as it requires a thermal diffusion time within the sample inside it. When a diffusion delay is reached, a data flow (data acquisitions) will then be visible between the MFIA and the measuring cell. The "Gh" pipe or line is blinking (alternating blue-grey, indicating a backflow from Gm to Gh). Here, we acknowledge the warning message by "OK", which sets its state to "ON". When the MFIA starts acquisitions, Z_at_T becomes green and we can see, in the "Measurements" tab, the progressive building of a Nyquist diagram characterizing the sample which is inside of the cell.

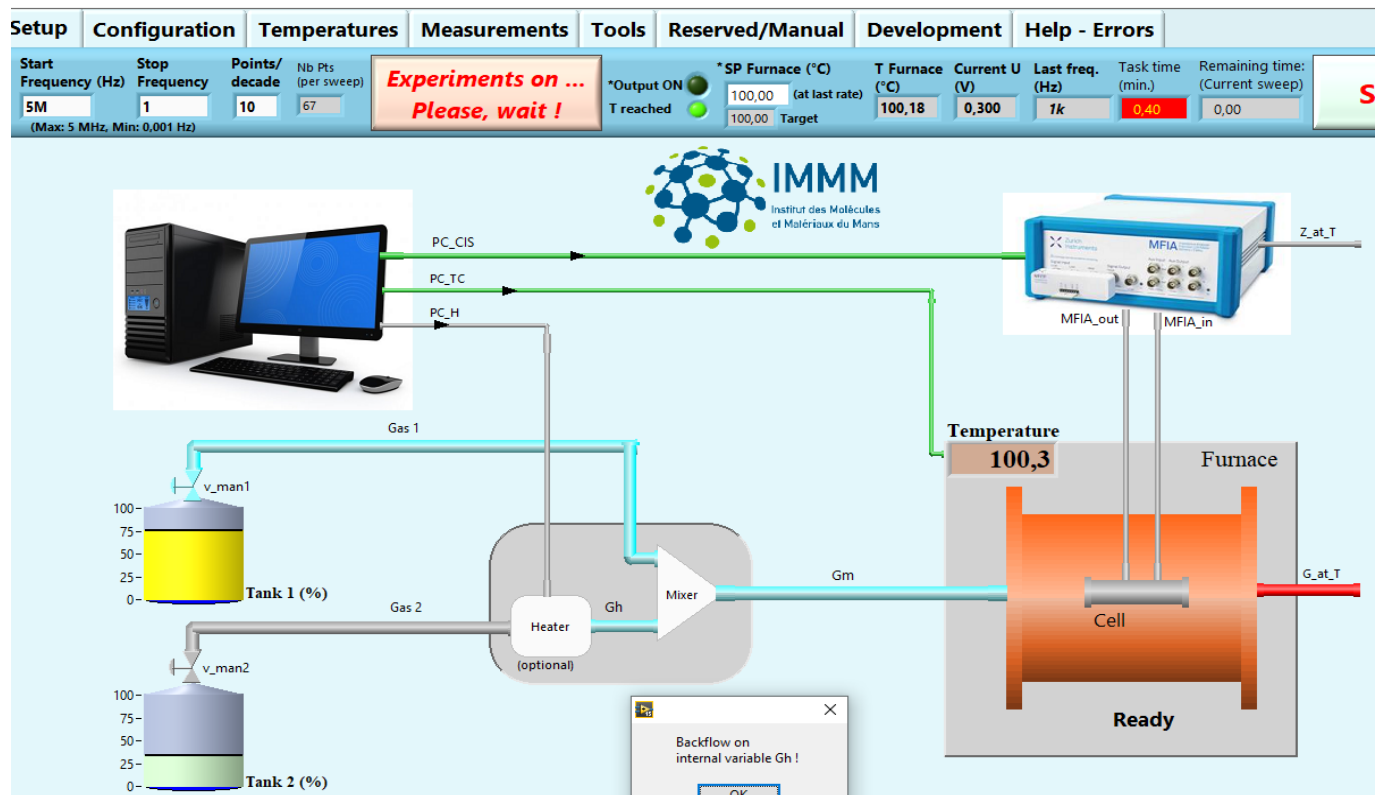

**Fig. 6.** Application of the methodology in a complex CIS system performing analysis of impedances (as a function of frequency) at different temperatures and with different mixed gases. Here, we have clicked on the manual valve "v_man1" (after opening it), then the supplied pipes/lines become colored (light blue) and we are warned of a possible backflow on Gh (this is deduced from step 3̄). The output G_at_T is red to indicate gas presence at the requested temperature (the thermocouple sensor is integrated in the furnace: after a diffusion delay, the sample is also at the requested temperature).

For this application, we reset (for example, every 20 iterations, depending on the application reaction time) the outputs and IVs (of the model) to "OFF" such that, if there are no more gases at the inputs, this simulates the progressive consumption of the gases contained in the IVs and outputs. If we had feedback values through sensors, we would, instead of this method, apply the real logical states. Without this initialization, there will never be a consumption of the gases or the data (consequently, some outputs would remain "ON", permanently). The "SP Furnace" value is already an active common parameter between the VS, the UM, and the PS ("Target" being the copy of this parameter from the PS). Therefore, if we change the value of "SP Furnace", the setpoint (on a temperature controller) in the PS will change accordingly, and vice-versa.

## 5. Conclusion

This article introduces a novel methodology for the modeling, simulation, control, and monitoring of physical systems or circuits through active Digital Twins. It also addresses bidirectional fluid flow challenges in circuits. The approach offers a systematic, active, and real-time solution, showcasing its applicability across diverse domains, including fluidic processes and electronic/electrical circuits. We argue and show that a behavior based Underlying Model of a Physical System is a necessary entity of any DT architecture. Based on this, we think that the DT concept is different from other known or emerging concepts, such as Machine learning, Internet of Things or the Network Communications. For instance, in this latter field/concept, some authors [17] have proposed a DT definition which is powerful for defaults analysis but it does not imply a two-way action between the DT and its twin (PS). A focus on the active or two-way DT concept, only, is needed in order to reach a clear and widely accepted definition of a DT.

Our methodology's easy integration into control systems and design software, coupled with its ability to detect abnormal situations, positions it as a useful tool for research and development. Through examples, the article illustrates its application to fluid distribution and electrical circuits. The detectable abnormal behaviors include the backflow of fluids (e.g., to avoid pollution) and potentials (e.g., to avoid electrocution), the leaks, and the blockages in physical circuits (especially the fluidic ones).

Looking ahead, further exploration and adaptation of this methodology to our evolving technological needs will enhance its utility. The systematic approach presented here establishes a foundation for future research and applications in the fast evolving landscape of Digital Twins.

### Declarations

**Funding** The author declares that no funds, grants, or other support were received during the preparation of this manuscript.

**Competing interests** The author has no relevant financial or non-financial interests to disclose.

**Author Contributions** The author assures the manuscript is not under consideration for publication and has not been published. The author is the only contributor to the study, conception, design and programming. All the previous versions of the manuscript were written by the author.

## Acknowledgment

The author acknowledges Dr. Romain Stomp, from Zurich Instruments (Zurich, Switzerland) for the useful information about the MFIA impedance analyzer control and Mr. Derek Johnson (Scribner Associates, Southern Pines, NC, USA), for the adaptation of impedance spectroscopy data formats supported by Zview analysis software (edited by Scribner Associates).